\documentclass[journal]{IEEEtran}
\usepackage[latin9]{inputenc}
\usepackage{amsmath}
\usepackage{amssymb}
\usepackage{graphicx}
\usepackage{esint}
\usepackage{verbatim}

\makeatletter

\usepackage{array}
\usepackage{url}
\usepackage{multirow}
\usepackage{cuted}
\usepackage{stfloats}
\usepackage[caption=false,font=normalsize,labelfont=sf,textfont=sf]{subfig}

\DeclareTextSymbolDefault{\textquotedbl}{T1}

\IEEEoverridecommandlockouts

\usepackage{cite}
\usepackage{amsfonts}
\usepackage{algorithmic}
\usepackage{textcomp}
\usepackage{xcolor}
\def\BibTeX{{\rm B\kern-.05em{\sc i\kern-.025em b}\kern-.08em
		T\kern-.1667em\lower.7ex\hbox{E}\kern-.125emX}}

\makeatother

\begin{document}
\title{Device-Free Localization with Multiple Antenna Receivers: Simulations
and Results \thanks{Funded by the European Union. Views and opinions expressed are however
those of the author(s) only and do not necessarily reflect those of
the European Union or European Innovation Council and SMEs Executive
Agency (EISMEA). Neither the European Union nor the granting authority
can be held responsible for them. Grant Agreement No: 101099491. Project Holden.}}

\author{
	\IEEEauthorblockN{Vittorio Rampa\IEEEauthorrefmark{1}, Federica Fieramosca\IEEEauthorrefmark{2}, Stefano Savazzi\IEEEauthorrefmark{1}, Michele D'Amico\IEEEauthorrefmark{2}}

\IEEEauthorblockA{\IEEEauthorrefmark{1} CNR-IEIIT, 
	Consiglio Nazionale delle Ricerche, Milan, Italy \\
	\{vittorio.rampa,stefano.savazzi\}@cnr.it}

\IEEEauthorblockA{\IEEEauthorrefmark{2} DEIB, 
	Politecnico di Milano, Milan, Italy \\
	\{federica.fieramosca,michele.damico\}@polimi.it}
	}

\pagenumbering{gobble}
 
\maketitle
\begin{abstract}
Device-Free Localization (DFL) is a passive radio method able to detect,
estimate, and localize targets (e.g., human or other obstacles) that
do not need to carry any electronic device. According to the \emph{Integrated
Sensing And Communication} (ISAC) paradigm, DFL networks exploit Radio
Frequency (RF) devices, used for communication purposes, to evaluate
also the excess attenuation due to targets moving in the monitored area, 
to estimate the target positions and movements. Several target models have
been discussed in the literature to evaluate the target positions by
exploiting the RF signals received by networked devices. Among these 
models, ElectroMagnetic (EM) body models emerged as an interesting research 
field for excess attenuation prediction using commercial RF devices. 
While these RF devices are usually single-antenna boards, 
the availability of low-cost multi-antenna devices e.g. those used 
in WLAN (Wireless Local Area Network) scenarios, allow us to exploit 
array-based signal processing techniques for DFL applications as well. 
Using an array-capable EM body model, this paper shows how to
employ array-based processing to improve angular detection
of targets. Unlike single-antenna devices that can provide only 
attenuation information, multi-antenna devices can provide
\emph{both} angular \emph{and} attenuation estimates about the target location. 
To this end, simulations are presented and preliminary results are discussed.
The proposed framework paves the way for a wider use of multi-antenna devices 
based, for instance, on WiFi6 and WiFi7 standards.
\end{abstract}


\section{Introduction}
\label{sec:intro} 
Device-Free Localization (DFL), a.k.a. passive
radio localization, is a transformative framework designed to detect,
localize, and track people (or objects) in a 3-D area illuminated
by Radio Frequency (RF) signals of opportunity. According to the \emph{Integrated
Sensing and Communication} (ISAC) paradigm~\cite{savazzi-2016},
the same RF devices used for communications can be leveraged by DFL
systems to transform each RF device into a \emph{virtual
sensor} for passive sensing operations. 

For instance, a network of these RF communication devices can be usefully
exploited to evaluate the presence and the location of people by using info from the received ElectroMagnetic (EM) field. In fact, both presence
and movements of people or objects, namely the \emph{targets}, modify
the incident EM field~\cite{koutatis-2010} that is already collected and processed by the RF devices to provide a reliable communication.
The alterations of the received RF signals does not only impair the radio communication channel, but can be also specifically processed
for localization purposes to estimate target information such as presence,
location, posture, movements, and size~\cite{savazzi-2016,shit-2019}.

Body models have been presented and widely discussed in the literature
for single-antenna devices for both single-target~\cite{wilson-2010,mohamed-2017,rampa-2017}
and multi-targets~\cite{rampa-2022} DFL scenarios. Moreover, different
radio channel measurements, such as Channel State Information (CSI),
Received Signal Strength (RSS), Angles of Arrival (AoA) and Time of
Flight (ToF)~\cite{savazzi-2016}, have been exploited for DFL applications
according to different processing frameworks such as radio imaging~\cite{wilson-2010},
Bayesian tracking~\cite{savazzi-2016}, fingerprinting methods~\cite{savazzi-2016},
Compressive Sensing algorithms~\cite{wang-2012}, and Machine Learning/Deep
Learning (ML/DL) systems~\cite{shit-2019}.

The concomitant wide diffusion of multi-antenna WLAN devices, such
as those designed according to the WiFi6 and WiFi7 standards, and
the development of CSI extraction tools~\cite{atif-2020}, have sped
up the research activities about multi-antenna DFL systems with the
adoption of multi-antenna CSI-based processing methods~\cite{shukri-2019,garcia-2020}.

A few references only focus on multi-antenna models:~\cite{ruan-2016}
employs a very simple propagation model while~\cite{ojeda-2022}
deals with a computational-intensive Ray Tracing (RT) approach not
suitable for real-time DFL applications. Actually, EM-based simulators
can be exploited for modeling as well. However, they are usually
too complex and very slow to be of practical use for real-time 
DFL~\cite{rampa-2017,rampa-2022}.

This paper extends the physical-statistical body model proposed by
the authors in~\cite{rampa-2022b,rampa-2023} to predict the body-induced
propagation losses using RF devices with multiple receiving antennas.
The array-based body model is exploited to simulate a target moving
in the monitored area and to provide an estimate of the AoA of the
perturbed RF waves due to the moving target.

The paper is organized as follows: in Sect.~\ref{sec:Array-based-body-model},
the array-based EM body model for applications with multiple receiving
antenna devices is briefly recalled. Then, Sect.~\ref{sec:impact_beamforming}
shows the array-based processing method to identify the AoA of the
RF signals that are altered by the presence of the target. Sect.~\ref{sec:simulation_results}
presents some preliminary simulation results while Sect.~\ref{sec:conclusions}
shows some preliminary conclusions and proposes future activities. 
\begin{center}
\begin{figure}[t]
\begin{centering}
\includegraphics[scale=0.35]{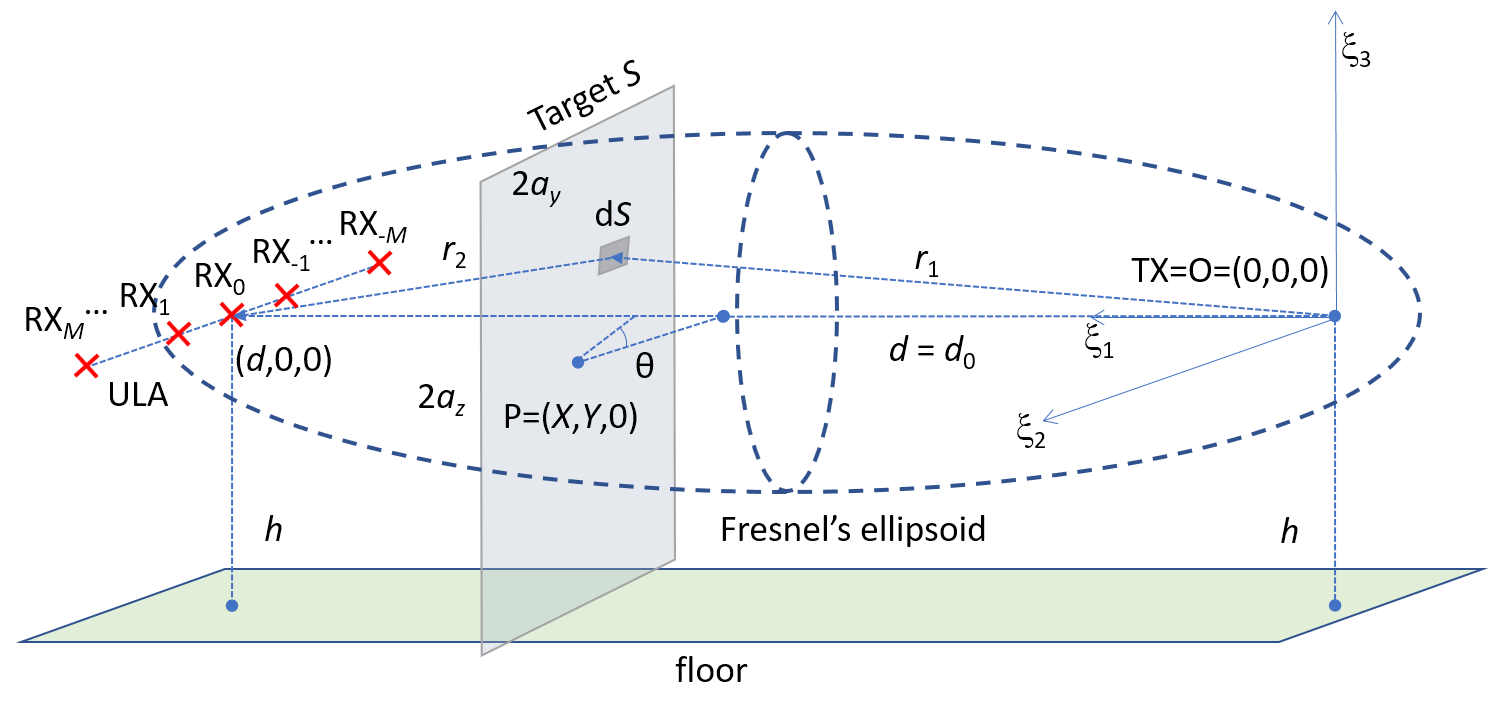} 
\par\end{centering}
\caption{Single-link 3-D layout of an ULA with $2M+1$ receiving antennas. The array is deployed along a line at distance $d=d_{0}$ from the TX and orthogonal to the LoS path connecting the $TX$ with the $RX_{0}$ device.}
\label{fig:array_layout} 
\end{figure}
\par\end{center}

\section{EM body model with multiple RX antennas}

\label{sec:Array-based-body-model}

According to the scalar diffraction framework adopted in~\cite{rampa-2017},
we briefly recall here the EM multi-antenna body model proposed 
in~\cite{rampa-2022b,rampa-2023} for a single-link scenario. 
Fig.~\ref{fig:array_layout} shows the layout composed by an Uniform 
Linear Array (ULA) of $2M+1$ isotropic receiver antennas RX$_{m}$, 
with$-M\leq m\leq M$ and $M,\,m\in\mathbb{Z}$,
a single isotropic transmitting antenna TX, and a single-target $S$
that moves in the monitored area near the radio link. Each \emph{m}-th
antenna RX$_{m}$ of the array is uniformly placed at mutual distance
$d_{a}$ along a segment orthogonal to the Line-of-Sight (LoS) at
distance $d=d_{0}$ from the TX and horizontally placed at distance
$h$ w.r.t. the floor. The central antenna is indicated by the index
$m=0$. We assume here that the floor has no influence on the RF signals. 
If this assumption does not hold true, the proposed model can easily be extended as shown in \cite{fieramosca-2023}. The 2-D footprint of the 3-D deployment of Fig.~\ref{fig:array_layout} is also depicted in Fig.~\ref{fig:array_layout-1}.

\subsection{Body model}

Considering DFL applications, we assume two main configurations:
the first one refers to the empty environment (i.e., the free-space case, namely $\mathcal{S}=0$) while, in the second scenario, the
target $S$ is present in the monitored area ($\mathcal{S}=1$). The
target is sketched~\cite{rampa-2017,rampa-2022} as a vertical standing EM absorbing 2-D sheet $S$ of height $2a_{z}$ and traversal
size $2a_{y}$ that is rotated of the angle $\theta$ w.r.t. the $\xi_{2}$
axis as shown in Figs.~\ref{fig:array_layout} and~\ref{fig:array_layout-1}.
Assuming a negligible mutual antenna coupling, approximately valid
for $d_{a}>\lambda/4$, the electric field $E^{\left(m\right)}$ received
by the \emph{m}-th antenna of the array, is:

\begin{equation}
\begin{array}{ll}
\frac{E^{\left(m\right)}}{E_{R}^{\left(m\right)}}= & 1-j\frac{d_{m}}{\lambda} {\displaystyle \intop_{S}} \frac{1}{r_{1,m}r_{2,m}}\cdot\\
 & \cdot\exp\left\{ -j\frac{2\pi}{\lambda}\bigl(r_{1,m}+r_{2,m}-d_{m}\bigr)\right\} d\xi_{2}\,d\xi_{3},
\end{array}\label{eq:E_full_array}
\end{equation}
where $E_{R}^{\left(m\right)}$ is the EM field received by the same $RX_{m}$ device in the reference condition $\mathcal{S}=0$. The term $d_{m}$ indicates
the distance of the \emph{m}-th antenna $RX_{m}$ of the array from
the TX while $d_{1,m}$ and $d_{2,m}$ are the distances of the projection
point $O_{m}^{'}$ of the barycenter $P$ of the 2-D surface $S$
from the TX and $RX_{m}$ nodes, respectively. Likewise, $r_{1,m}$ and $r_{2,m}$
are the distances of the generic elementary area $dS$ of the target
$S$ from the TX and the $RX_{m}$, respectively. Integration 
is performed in (\ref{eq:E_full_array}) over the squared domain
$S$ having height $2a_{z}$ and traversal size $2a_{y}$. $P(X,Y,Z)$ indicates  
the target location coordinates. For simplicity, $Z$ is then dropped since it is implicitly assumed $Z=0$.

For $M=0$, the equation (\ref{eq:E_full_array}) reduces to the single-antenna
case~\cite{rampa-2017} where RX$_{0}$ coincides with the RX antenna
at distance $d=d_{0}$ from the TX. The excess attenuation~\cite{rampa-2017}
at RX$_{m}$, due to the influence of the target $S$ w.r.t. the free-space
scenario, is computed~\cite{rampa-2017} as $A_{T}^{\left(m\right)}=P_{R}^{\left(m\right)}/P^{\left(m\right)}=\left|E_{R}^{\left(m\right)}/E^{\left(m\right)}\right|^{2}$
where $P_{R}^{\left(m\right)}$ and $P^{\left(m\right)}$ are the
received power at RX$_{m}$ without and with a target in the link
area, respectively. Usually, the excess attenuation is given in dB
as $A_{T,dB}^{\left(m\right)}=10\,\log_{10}\left|E_{R}^{\left(m\right)}/E^{\left(m\right)}\right|^{2}$.
The ratio $E_{R}^{\left(m\right)}/E_{R}^{\left(0\right)}$
is given by:

\begin{equation}
\begin{array}{ll}
\frac{E_{R}^{\left(m\right)}}{E_{R}^{\left(0\right)}}= & \frac{d_{0}}{d_{m}}\,\exp\left\{ -j\frac{2\pi}{\lambda}\bigl(d_{m}-d_{0}\bigr)\right\} ,\end{array}\label{eq:E_R_array_0}
\end{equation}
where $E_{R}^{\left(0\right)}$ is the electric field received by
the central antenna $RX_{0}$ that is on the LoS path at distance
$d_{0}$ from the TX. Eq. (\ref{eq:E_full_array}) can be rewritten as:

\begin{equation}
\begin{array}{ll}
\frac{E^{\left(m\right)}}{E_{R}^{\left(0\right)}}= & \frac{d_{0}}{d_{m}}\,\exp\left\{ -j\frac{2\pi}{\lambda}\left(d_{m}-d_{0}\right)\right\} \bigl[1-j\frac{d_{m}}{\lambda}{\displaystyle \intop_{S}}\frac{1}{r_{1,m}r_{2,m}}\cdot\bigr.\\
 & \bigl.\cdot\exp\left\{ -j\frac{2\pi}{\lambda}\left(r_{1,m}+r_{2,m}-d_{m}\right)\right\} d\xi_{2}\,d\xi_{3}\bigr].
\end{array}\label{eq:E_full_array_new}
\end{equation}

\begin{center}
\begin{figure}[t]
\begin{centering}
\includegraphics[scale=0.56]{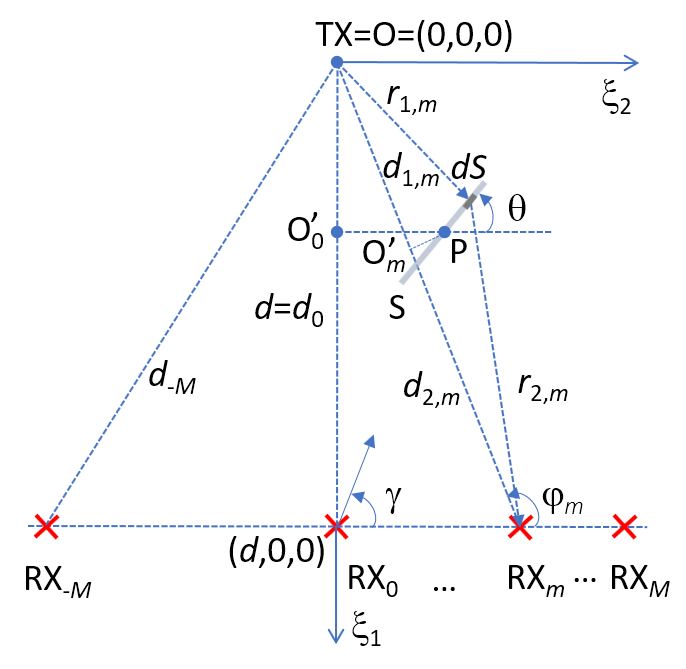}
\par\end{centering}
\caption{2-D layout of the radio link of Fig.~\ref{fig:array_layout}. The
point $O_{m}^{'}$ is the projection of the barycenter $P$ of the
target $S$ over the \emph{m}-th LoS path having length $d_{m}$.}\label{fig:array_layout-1}
\end{figure}
\par\end{center}

\noindent According to (\ref{eq:E_R_array_0}) and (\ref{eq:E_full_array_new}),
the \emph{m}-th component of the received vector $\mathbf{r}\left(t,\mathcal{S}\right)=\left[r_{-M}\left(t,\mathcal{S}\right)\,...\,r_{-1}\left(t,\mathcal{S}\right)\,r_{0}\left(t,\mathcal{S}\right)\,r_{1}\left(t,\mathcal{S}\right)\,...\,r_{M}\left(t,\mathcal{S}\right)\right]^{T}$
of size $2M+1$ is defined as:

\noindent
\begin{equation}
r_{m}\left(t,\mathcal{S}\right)=\left\{ \begin{array}{ll}
E_{R}^{\left(m\right)}+n_{m}=E_{R}^{\left(0\right)}\,\frac{E_{R}^{\left(m\right)}}{E_{R}^{\left(0\right)}}+n_{m} & \textrm{if \ensuremath{\mathcal{S}=0}}\\
E^{\left(m\right)}+n_{m}=E_{R}^{\left(0\right)}\,\frac{E_{R}^{\left(m\right)}}{E_{R}^{\left(0\right)}}\,\frac{E^{\left(m\right)}}{E_{R}^{\left(m\right)}}+n_{m} & \textrm{if \ensuremath{\mathcal{S}=1}}
\end{array}\right.\label{eq:signal_model_reduced}
\end{equation}
where $n_{m}$ is the \emph{m}-th component of the Additive White
Gaussian Noise (AWGN) complex vector $\mathbf{n}\left(t\right)=\left[n_{-M}\left(t\right)\,...\,n_{-1}\left(t\right)\,n_{0}\left(t\right)\,n_{1}\left(t\right)\,...\,n_{M}\left(t\right)\right]^{T}$
of size $2M+1$, that is assumed to be spatially white with zero mean
and covariance $\sigma_{n}^{2}\mathbf{I}$. 

The interested reader can refer to~\cite{savazzi-2016,rampa-2017,rampa-2022} for further information and discussions about the noise effects with and without the presence of a target inside the monitored area.

\subsection{Beamforming}

Being $\boldsymbol{w}=\left[w_{-M}\,...\,w_{0}\,...\,w_{M}\right]^{T}$ the
beamforming vector of size $2M+1$ collecting all linear beamforming
coefficients, the output $y=y\left(t,\mathcal{S}\right)$ of linear
beamforming processing is equal to:

\begin{equation}
y\left(t,\mathcal{S}\right)=\sum_{m=-M}^{+M}w_{m}^{*}\,r_{m}\left(t,\mathcal{S}\right)=\mathbf{w}^{H}\mathbf{r}\left(t,\mathcal{S}\right),\label{eq:beam_forming}
\end{equation}

\noindent where $\left(\cdot\right)^{*}$ indicates the conjugate
operator and $\left(\cdot\right)^{H}$ the Hermitian operator, while
$r_{m}\left(t,\mathcal{S}\right)$ is the EM field received in the
$RX_{m}$ antenna. 

In DFL scenarios, we are interested in evaluating
the power $P_{y}\left(\mathcal{S}\right)$ of the signal $y$ with
or without the target $S$. This is given by:

\begin{equation}
\begin{array}{ll}
P_{y}\left(\mathcal{S}\right) & =\mathbb{E}\left[y\left(t,\mathcal{S}\right)\,y^{*}\left(t,\mathcal{S}\right)\right]\\
 & =\mathbf{w}^{H}\,\mathbf{R}\left(\mathcal{S}\right)\,\mathbf{w}\,
\end{array}\label{eq:beam_forming_power}
\end{equation}
where matrix $\mathbf{R}\left(\mathcal{S}\right)=\mathbb{E}\left[\mathbf{r}\left(t,\mathcal{S}\right)\,\mathbf{r}^{H}\left(t,\mathcal{S}\right)\right]$
of size $\left(2M+1\right)\times\left(2M+1\right)$ is the autocorrelation
matrix of $\mathbf{r}\left(t,\mathcal{S}\right))$.

\begin{figure}
\begin{centering}
\includegraphics[scale=0.54]{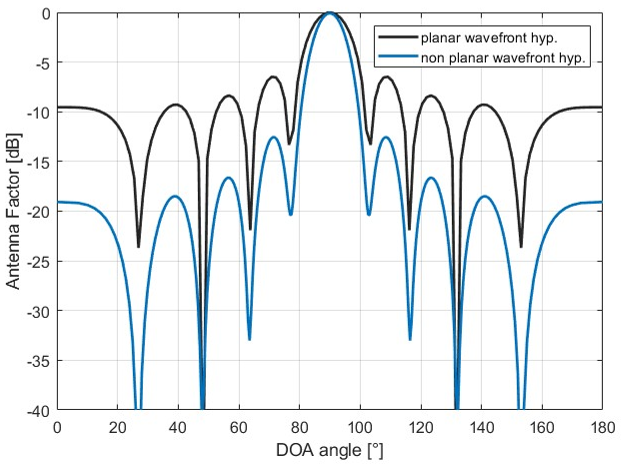}
\par\end{centering}
\caption{Comparisons of the array factors for an ULA composed by $9$ antennas
($M=4$) uniformly spaced at $d_{a}=\lambda/2$ with planar and
non planar wave propagation hypotheses. For the non planar assumption,
it is also $d_{0}=d$ and $d_m=\sqrt{d_0^2+m^2 d_a^2}$ with $d=4$ m. }
\label{fig:array_factor}
\end{figure}

\section{Impact of body effects on indoor beamforming}
\label{sec:impact_beamforming}

The presence of the target generates perturbations in the beamforming response, occasionally altering the direction of maximum received radiation. This effect is analyzed in this section because of its relevance to passive localization applications. 

In addition to discussing the effects of body motions on indoor beamforming response, we also propose an estimation algorithm to evaluate the direction of arrival (DoA) $\gamma$ of the electromagnetic wave when a  target is present on the scene. To this aim, (\ref{eq:signal_model_reduced}) can be rewritten as:
\begin{equation}
\mathbf{r}\left(t,\mathcal{S}\right)=\left\{ \begin{array}{ll}
E_{R}^{\left(0\right)}\,\mathbf{a}+\mathbf{n} & \textrm{if \ensuremath{\mathcal{S}=0}}\\
E_{R}^{\left(0\right)}\,\textrm{diag\ensuremath{\left(\mathbf{a}\right)}}\,\mathbf{E}_{r}+\mathbf{n}= & \textrm{if \ensuremath{\mathcal{S}=1},}
\end{array}\right.\label{eq:signal_model}
\end{equation}
where the column vector $\mathbf{E}_{r}$ of size $2M+1$ represents the electric field ratio (\ref{eq:E_full_array}) received by the antenna array for $\ensuremath{\mathcal{S}=1}$. It is defined as $\mathbf{E}_{r}=\left[E^{\left(-M\right)}/E_{R}^{\left(-M\right)} \, ... \, E^{\left(M\right)}/E_{R}^{\left(M\right)}\right]^T$.

The steering vector $\mathbf{a}$ depends on the propagation assumptions:
for planar wave propagation, it is, for $-M\leq m\leq M$:
\begin{equation}
\mathbf{a}=\left[\exp\left\{ j\,m\,\frac{2\pi}{\lambda}d_{a}\cos\gamma\right\} \right].
\label{eq:planar_wavefront}
\end{equation}
Since DFL applications
are mostly deployed in indoor scenarios with short links, the planar wavefront assumption~\cite{benesty-2021} adopted for far-field scenarios is an approximation of the true behavior. Thus, the correct steering vector is equal to~\cite{rampa-2022}: 
\begin{equation}
\mathbf{a}=\left[\frac{d_{0}}{d_{m}}\,\exp\left\{ j\,m\,\frac{2\pi}{\lambda}d_{a}\frac{\cos\left\{ \left(\gamma+\varphi_{m}\right)/2\right\} }{\cos\left\{ \left(\gamma-\varphi_{m}\right)/2\right\} }\right\} \right],
\label{eq:a_true}
\end{equation}
where $\varphi_{m}$ is the angle formed by the LoS of the \emph{m}-th
array antenna with the $\xi_{2}$ axis. For $-M\leq m\leq M$, it
is $\varphi_{m}=\arcsin\left\{ \left(d_{0}/d_{m}\right)\,\sin\gamma\right\} $.
Moreover, it is also $\left|\mathbf{a}\right|^{2}=\sum_{m=-M}^{M}\left(d_{0}/d_{m}\right)^{2}$
and $\varphi_{0}=\gamma$. 

The array factor $F_{a}=\mathbf{w}^{T}\mathbf{a}\left(\gamma\right)$
defines the response of the array as a function of the selected beamforming
coefficients $\mathbf{w}$ and of the incident angle $\gamma$ of
the steering vector $\mathbf{a}\left(\gamma\right)$. According to
the selected outdoor (i.e., long or very long LoS path) or indoor 
(i.e., short LoS path) scenarios, corresponding to planar vs non-planar hypotheses,
respectively, (\ref{eq:array_factor}) and
Fig.~\ref{fig:array_factor} summarize the analytical and graphical assumptions.

\begin{equation}
F_{a}=\left\{ \begin{array}{ll}
\frac{1}{2M+1} {\displaystyle \sum_{m=-M}^{+M}} \exp\{{jm\frac{2\pi}{\lambda}d_{a}\cos\gamma}\} & \hskip -2mm \textrm{planar hyp.}\\
\frac{1}{2M+1} {\displaystyle \sum_{m=-M}^{+M}} \frac{d_{0}}{d_{m}}\, \cdot \\ \cdot \exp\left\{ j\,m\,\frac{2\pi}{\lambda}d_{a}\frac{\cos\left\{ \left(\gamma+\varphi_{m}\right)/2\right\} }{\cos\left\{ \left(\gamma-\varphi_{m}\right)/2\right\} }\right\}  & \hskip -2mm \textrm{non-planar hyp.}
\end{array}\right.
\label{eq:array_factor}
\end{equation}

However, to exploit (\ref{eq:a_true}), all distances $d_{m}$ must be a priori known. For this reason, in this paper, we assume as a starting hypothesis the approximation (\ref{eq:planar_wavefront}) for $\mathbf{w}$. Other assumptions and methods can be adopted, but these are outside the scope of this paper.

As assumed for the single-antenna case~\cite{rampa-2017}, but differently from what suggested in~\cite{rampa-2022}, in this paper we assume that the mean
excess attenuation $A_{T}$ corresponds to the ratio
between the power terms $P_0\left(\mathcal{S}=0\right)=\mathbb{E}\left[ r_0\left(t,\mathcal{S}=0\right)r_0^{*}\left(t,\mathcal{S}=0\right)\right]$ and $P_{y}\left(\mathcal{S}=1\right)$. $P_0\left(\mathcal{S}=0\right)$ is the power received by the central antenna of the array (i.e., $m=0$) when there are no targets within the monitored area. 

By neglecting the noise terms, the excess attenuation is given by
$A_{T,dB}= 10\log_{10}\frac{P_{0}\left(\mathcal{S}=0\right)}{P_{y}\left(\mathcal{S}=1\right)}$.

To locate the target, we estimate the angle of arrival $\hat{\gamma}$, namely the angle/direction of maximum received power w.r.t the empty environment, with no subjects in the monitored area:
\begin{equation}
\hat{\gamma}=\arg\max_{0 \le\gamma\le\pi}\frac{P_{y}\left(\mathcal{S}=1\right)}{P_{0}\left(\mathcal{S}=0\right)}
\label{fig: maximization}
\end{equation}
that maximizes the power ratio $P_y/P_0$ by using (\ref{eq:signal_model}) and the steering vector $\mathbf{a}\left(\gamma\right)$ given by (\ref{eq:planar_wavefront}). Then, the corresponding value of $A_{T,dB}=A_{T,dB}\left(\hat{\gamma}\right)$ is evaluated, as well.

With respect to~\cite{rampa-2017}, the use of an antenna array allows us to estimate two target information: the excess attenuation due to the target and the DoA of the received signals that are distorted by the target. It is worth noticing that the DoA of the received signals is influenced by the target but does not coincide with the one due to the location of the target.

\begin{figure}[t]
\begin{centering}
\includegraphics[scale=0.63]{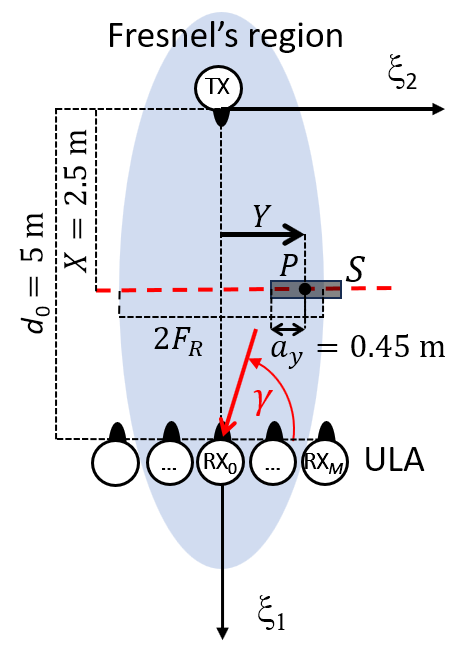} 
\par\end{centering}
\caption{Link layout used for the simulations: the antenna array is composed
by 5 antennas ($M=2$) while the target $S$, with size $a_{z}=0.9$
m and $a_{y}=0.45$ m, is placed in different positions along the
line having distance $x=2.5$ m from the TX.}\label{fig:layout5}
\end{figure}

\section{Simulation results}
\label{sec:simulation_results}

In this section, we introduce the simulation setup and show some preliminary results obtained according to the criterion (\ref{fig: maximization}).

The carrier frequency $f_c$ is set to $f_c=2.4868$ GHz (i.e., $\lambda=12$ cm), and we used an ULA composed of $5$ omni-directional antennas ($M=2$) uniformly spaced at $d_a=\lambda/2$ as shown in the layout sketch described in Fig.~\ref{fig:layout5}. The length of the central link of the array (i.e., for $m=0$) is equal to $d=d_0=5$ m while all links of the array are horizontally placed at height $h=0.9$ m from the ground. The minor semi-axis of the first Fresnel's ellipsoid $F_R= \sqrt{\lambda\,d}/2$ is equal to $F_R=0.39$ m. We also assume that there are no reflections or other multi-path effects due to the presence of floor, walls, and ceiling.

As far as the body model is concerned, the absorbing 2-D sheet that represents the target has size $a_z=0.9$ m and $a_y=0.45$ m (i.e. corresponding to a total size of $1.80$ m $\times$ $0.90$ m). The target is placed vertically on the floor, that is used only to support the target and does not have any EM influence on the radio links. The LoS of the central link will be the reference LoS line for the target positions with the origin of the axes placed in the TX as shown in Fig.~\ref{fig:layout5}.

\begin{figure}[t]
\begin{centering}
\includegraphics[scale=0.335]{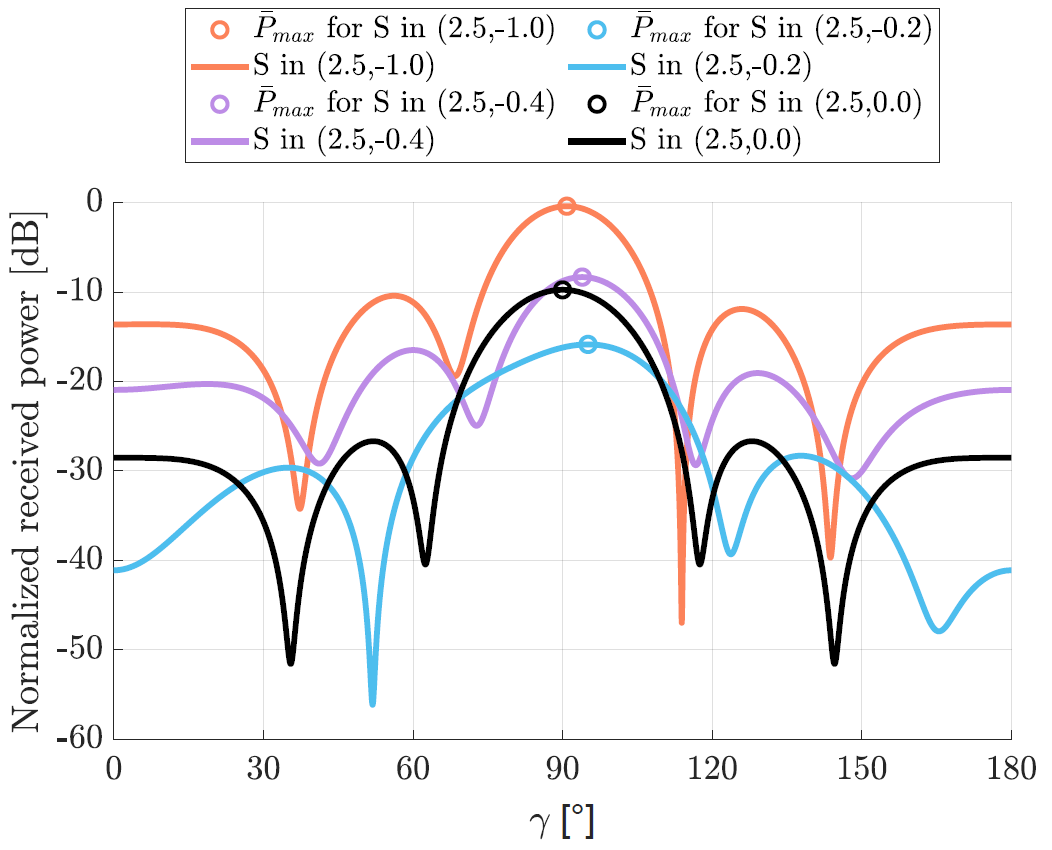}  
\par\end{centering}
\caption{The received power ratio $P_{y}\left(\mathcal{S}=1\right)/P_{0}\left(\mathcal{S}=0\right)$ as a function of $\gamma$ and different target positions: the orange line indicates a target in (2.5,-1.0), the purple line in (2.5,-0.4), the blue line in (2.5,-0.2), and the black line in (2.5,0.0). The small circles indicate the maximum values of the received power ratio.}\label{fig:left}
\end{figure}

\begin{figure}[t]
\begin{centering}
\includegraphics[scale=0.34]{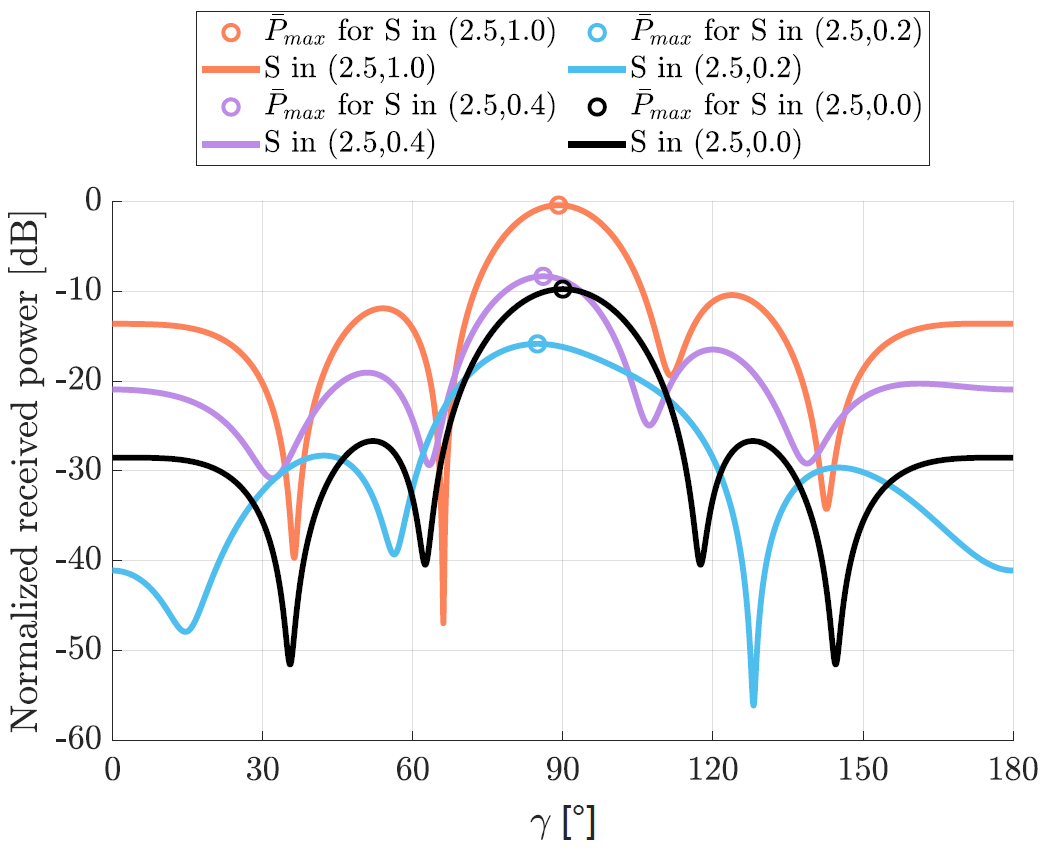} 
\par\end{centering}
\caption{The received power ratio $P_{y}\left(\mathcal{S}=1\right)/P_{0}\left(\mathcal{S}=0\right)$ as a function of $\gamma$ and different target positions: the orange line indicates a target in (2.5,1.0), the purple line in (2.5,0.4), the blue line in (2.5,0.2), and the black line in (2.5,0.0). The small circles indicate the maximum values of the received power ratio.}\label{fig:right}
\end{figure}

It is well-known~\cite{rampa-2017,rampa-2022} that a target influences the received field only when it is near the radio link. The same behavior is apparent also here as shown in Figs.~\ref{fig:left} and \ref{fig:right} where the received power ratio $P_{y}\left(\mathcal{S}=1\right)/P_{0}\left(\mathcal{S}=0\right)$ is depicted as a function of $\gamma$.

It is apparent that the proposed method is capable of discriminating between targets placed on the left side of the link i.e. for $\pi/2 < \gamma \le \pi$ (Fig.~\ref{fig:left}) or right side of the link i.e. for $0 \le \gamma < \pi/2$  (Fig.~\ref{fig:right}) at least for an interval around $(\pi/2,0)$ in the $\gamma-Y$ space. This capability cannot be implemented for a single-antenna receiver that shows a perfect symmetry of the received power ratio w.r.t. the LoS path~\cite{rampa-2017}.

\begin{figure}[t]
\begin{centering}
\includegraphics[scale=0.295]{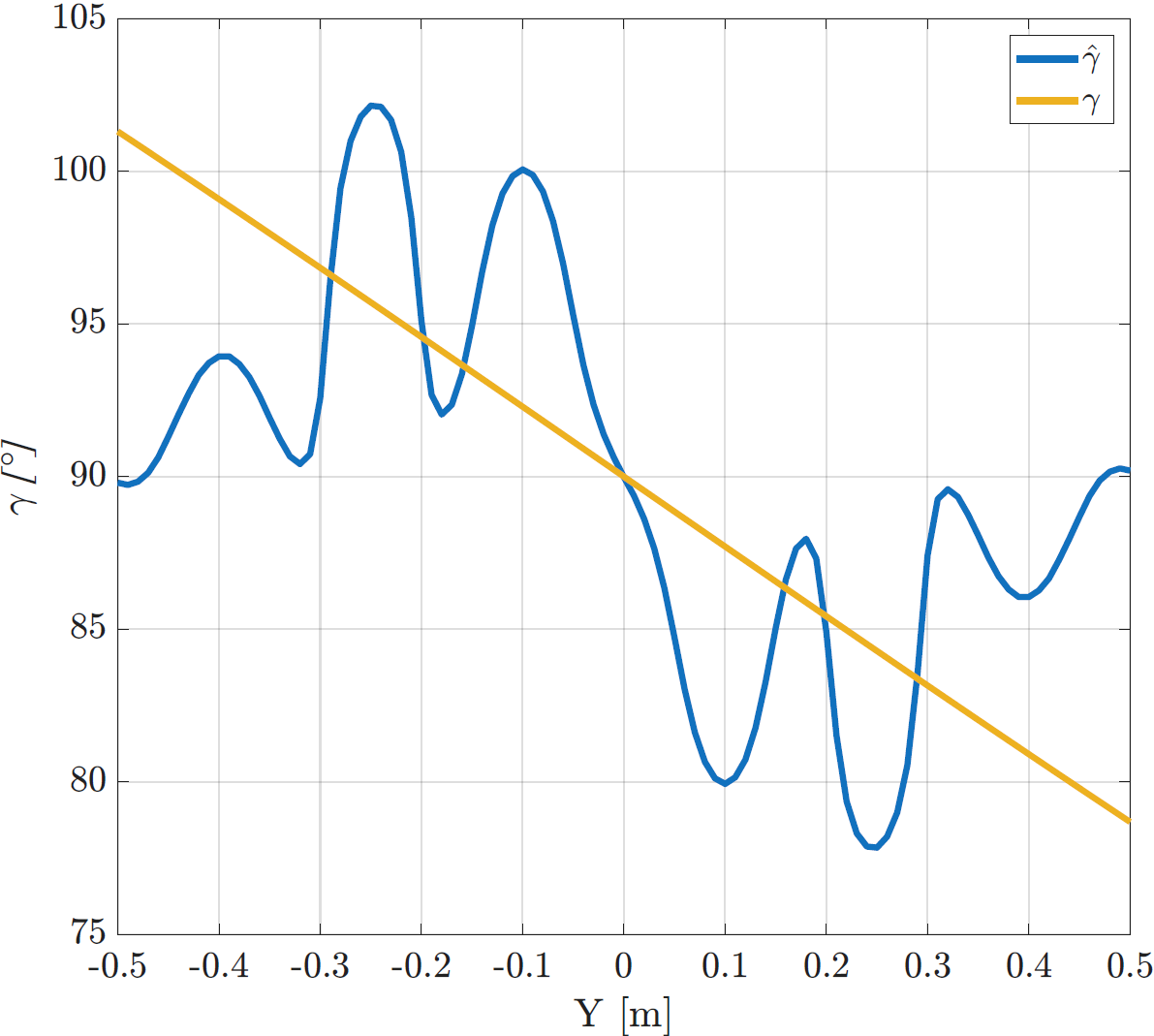} 
\par\end{centering}
\caption{Estimated ($\hat{\gamma}$) vs true ($\gamma$) values of the angle of arrival for different target positions located in (2.5,Y) with varying Y.}\label{fig:comparison}
\end{figure}

Fig.~\ref{fig:comparison} shows the comparison between the true $\gamma$ and the estimated $\hat{\gamma}$ values for target positions close to the link. It is worth noticing that, for this example, the size of the Fresnel's ellipsoid is the Y interval $[-0.39, +0.39]$ m and the side discrimination capability interval is approximately the same. 

\begin{figure}[t]
\begin{centering}
\includegraphics[scale=0.41]{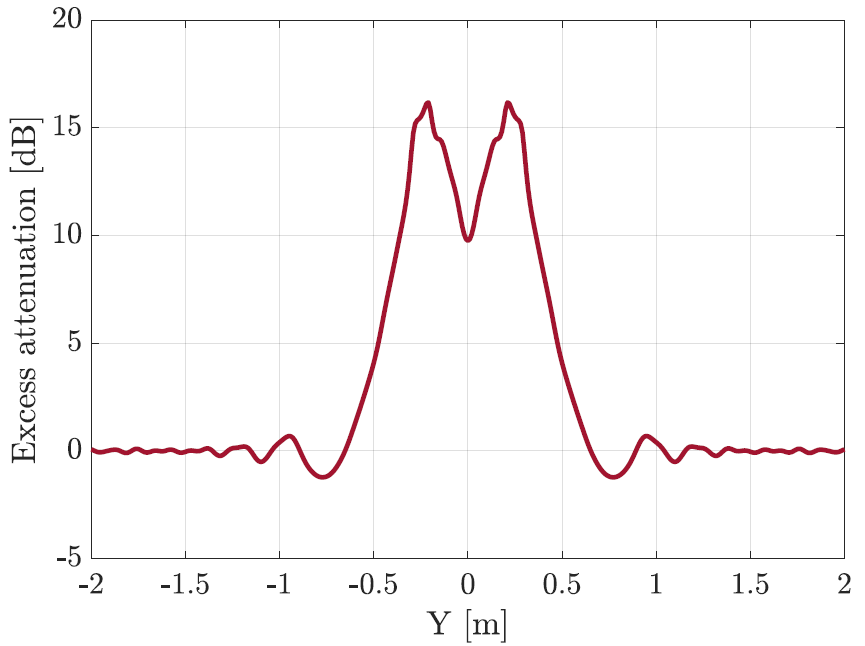}
\par\end{centering}
\caption{Excess attenuation for the different target positions located in (2.5,Y) with varying Y. These values are obtained for the $\hat{\gamma}$ values of Fig.~\ref{fig:comparison}. To show the slow damping of the excess attenuation, the Y interval is larger than the one of Fig.~\ref{fig:comparison}.}
\label{fig:excess_attenuation}
\end{figure}

Fig.~\ref{fig:excess_attenuation} shows the excess attenuation in dB corresponding to the $\hat{\gamma}$ values shown in Fig.~\ref{fig:comparison}. It has to be noted that this side discrimination capability can be exploited only for target close to the radio links since these links are not influenced by distant target as apparent in Fig.~\ref{fig:excess_attenuation} due to very small values of the excess attenuation for $|Y|>1$ m.

\section{Conclusions}
\label{sec:conclusions}  

In this paper, we introduce a body model for linear antenna arrays capable of inferring both the presence of a target in the surrounding of the radio link, and also the angular information, namely the Direction of Arrival (DoA). In fact, the received signals that are distorted by the target, reveal the offset position of the same target w.r.t. the array line-of-sight path. The model is relevant for Device-Free Localization (DFL) applications.

Unlike single-antenna links that can only evaluate body-induced attenuation effects in the monitored area covered by the radio links, the multi-antenna array is able to evaluate both attenuation and Angle-of-Arrival (AoA) information. 

The presence of the target deforms the received ElectroMagnetic (EM) field w.r.t. the free-space configuration and this distortion can be used to estimate angular positioning information. This AoA-related capability paves the way to the use of array-based sensing methods for more accurate passive localization as well as object detection systems.

\end{document}